\documentclass[compsoc,conference,letter,10pt,times]{IEEEtran}
\IEEEoverridecommandlockouts
\usepackage{amsmath,amssymb,amsfonts}
\usepackage{algorithmic}
\usepackage{graphicx}
\usepackage{textcomp}
\usepackage{bmpsize}
\usepackage{xcolor}
\usepackage{lipsum}
\usepackage[colorlinks=true,urlcolor=black]{hyperref}
\def\BibTeX{{\rm B\kern-.05em{\sc i\kern-.025em b}\kern-.08em
		T\kern-.1667em\lower.7ex\hbox{E}\kern-.125emX}}
\usepackage{comment}
\usepackage{subcaption}
\begin{document}
	
	\title{The Impact of Visibility on the Right to Opt-out of Sale under CCPA}
	
	\author{\IEEEauthorblockN{Aden Siebel}
		\IEEEauthorblockA{\textit{Pomona College} \\
			amsa2017@mymail.pomona.edu}
		\and
		\IEEEauthorblockN{Eleanor Birrell}
		\IEEEauthorblockA{\textit{Pomona College}\\
			eleanor.birrell@pomona.edu}
	}
	
	
	\maketitle
	
	\begin{abstract}
		The California Consumer Protection Act (CCPA) gives users the right to opt-out of sale of their personal information, but prior work has found that opt-out mechanisms provided under this law result in very low opt-out rates. Privacy signals offer a solution for users who are willing to proactively take steps to enable privacy-enhancing tools, but many users are not aware of their rights under CCPA. We therefore explore an alternative approach to enhancing privacy under CCPA: increasing the visibility of opt-out of sale mechanisms. We conduct an user study with 54 participants and find that visible, standardized banners significantly increase  opt-out of sale rates in the wild. Participants also report less difficulty opting out and more satisfaction with opt-out mechanisms compared to the native mechanisms currently provided by websites. Our results suggest that effective privacy regulation depends on imposing clear, enforceable visibility standards, and that CCPA's requirements for opt-out of sale mechanisms fall short. 
	\end{abstract}
	
	
\section{Introduction}

A rise in awareness of how data are collected and used online has resulted in several recent Internet privacy and data protection regulations (e.g., GDPR~\cite{gdpr}, CCPA~\cite{ccpa}). 
One of the rights granted to Californians under CCPA is the right to opt-out of sale of their personal information. However, prior work suggests that current implementations of opt-out mechanisms fail to effectively empower users to opt-out of sale: most websites provide this link in a small font at the bottom of the page (often only accessible after scrolling)~\cite{oconnor2021clear}, and these links are rarely noticed by users~\cite{cranor2021informing}.

Privacy signals, such as Global Privacy Control (GPC)~\cite{GPC}, offer one approach to enhancing privacy by enabling users to universally signal a desire to opt-out of sale, and current guidelines issued by the California Attorney General require websites to respect these opt-out signals. However, GPC is currently only supported by browsers with market share under 4\% (e.g., Firefox, Brave, Duck Duck Go) and via Chrome browser-extensions (e.g., OptMeowt~\cite{Zimmeck20}). Moreover, past experience with privacy signals like Do Not Track (DNT) suggests that GPC will only be considered a valid opt-out signal as long as it is turned off by default.   GPC therefore is---and is likely to remain---only a solution for people who (1) are aware of their right to opt-out of sale and (2) take specific, proactive steps to invoke their right to opt-out by enabling GPC. Since many Californians are still unaware of their rights under CCPA~\cite{cpi2022survey}, opt-in privacy signals alone are insufficient to empower all users with the right to opt-out of sale.


This work investigates an alternate approach to enhancing privacy under CCPA: improving the visibility of opt-out mechanisms. Unlike prior work, we evaluated the impact of improved visibility on real-world behavior. To do so, we designed and implemented two browser extensions: (1) an observational extension that automatically detects whether a website sells personal information and records whether users invoke their right to opt out and (2) CCPA Opt-out Assistant (COA), a browser extension that extends the observational extension by providing a simple, standardized banner that links to existing opt-out mechanisms. We then used these two extensions to evaluate the impact of  visibility on opt-out rates. We conducted an observational user study ($n=54$) 
in which we recorded real-world user opt-out behavior with these two extensions over a period of one month.  

We found that the presence of banner-based opt-out of sale mechanisms significantly increased engagement. On average, real-world COA users opted-out of 18.8\% of websites that provided an opt-out of sale link, and the majority of opt-outs used mechanisms provided by COA rather than the links provided directly by websites.  COA users were also less likely to describe the opt-out process as difficult or to be unsatisfied with available opt-out mechanisms. These results suggest that enhancing the visibility of opt-out mechanisms would significantly improve privacy under CCPA.

While we have now made COA publicly available on the Chrome store, we do not consider this extension itself to be an effective tool for enhancing privacy under CCPA; in fact, extensions and browser settings that support privacy signals are probably a better solution for users who are willing and able to proactively take steps to invoke their rights under CCPA. Instead, we view our results as evidence that effectively extending privacy rights to all users depends on regulations imposing minimum visibility standards, and that CCPA's requirement of a clear and conspicuous link on the home page falls short of this standard. Future regulations will need to provide clear and enforceable visibility requirements informed by empirical user studies in order to ensure that they actually enhance user privacy.

\section{Background and Related Work}

The primary goal of the CCPA is to give users more control over their personal information. It therefore introduced four key rights:
\begin{enumerate}
	\item \textbf{The right to know.} Users have a right to know what personal information a business collects and how that information is used and shared. 
	\item \textbf{The right to delete.} Users have a right to delete personal information about them (with some exceptions).
	\item \textbf{The right to opt-out of sale.} Users have the right to opt-out of the sale of their personal information. Businesses must provide a ``a clear and conspicuous link'' on the homepage of their website entitled ``Do Not Sell My Personal Information'' that enables users to invoke this right. 
	\item \textbf{The right to non-discrimination.} Businesses cannot deny a service, degrade the quality of service, or change the price of a service as retaliation for exercising these rights. 
\end{enumerate}
CCPA also broadened the definition of personal information; CCPA's  
definition explicitly includes online activities (e.g., interactions with a website) and any inferences drawn from such information.

Prior work has investigated how websites implement CCPA's opt-out of sale requirement, and how those implementation choices impact user privacy. O'Connor et al.~\cite{oconnor2021clear} found that just 7.8\% of websites that sell data provide an opt-out link in a banner; most websites (80.9\%) provide an opt-out of sale link somewhere on the homepage, but that link ofter requires the user to scroll to the bottom of the page or to interact with clickable elements before it is visible. These links are rarely noticed~\cite{cranor2021informing} and difficult to find~\cite{mahoney2020california}, and they result in lower engagement~\cite{oconnor2021clear} and frequent failed opt-out attempts~\cite{mahoney2020california} in experimental settings. These results are consistent with  work on other opt-out mechanisms, which are also hard for users to  understand and use~\cite{habib2020s,habib2019empirical,Leon12,Utz19,sanchez2019can,nouwens2020dark,machuletz2020multiple,sakamoto2019after}.

Several approaches to enhancing the right to opt-out of sale have been proposed. One approach is to issue standardized privacy signals~\cite{GPC,OptOut,Zimmeck20}; however, privacy signals only enhance privacy for users who are aware of their rights and take steps to enable the signal. CCPA Detector~\cite{CCPAdetector} is an extension that detects CCPA-related privacy policies; however it is unable to accurately detect which websites sell data and its effect on end-user privacy has not been evaluated. Cranor et. al.~\cite{cranor2020design,cranor2020user,cranor2020ccpa,cranor2021informing} examined how different taglines and icons influence user comprehension and recall of Do Not Sell links; they recommended the adoption of standardized icons and placement. However, their work did not compare the usability of links or privacy icons with opt-out banners.

 \begin{figure*}[t!]
 	\begin{center}
 		
 		\begin{subfigure}{.3\textwidth}
 			\includegraphics[width=\columnwidth]{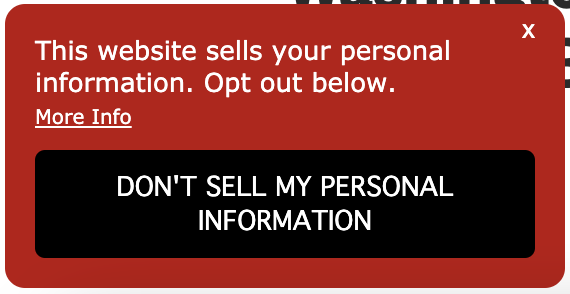}
 			\caption{Banner for  a valid link.}
 			\label{fig:popup-closeup}
 		\end{subfigure}
 		\hspace{10pt}
 		\begin{subfigure}{.3\textwidth}
 			\includegraphics[width=\columnwidth]{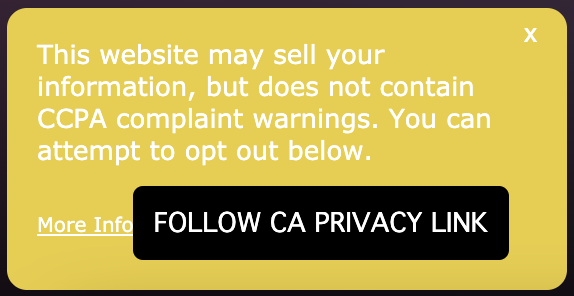}
 			\caption{Banner for an ambiguous link.}
 			\label{fig:warning-closeup}
 		\end{subfigure}
 		\hspace{10pt}
 		\begin{subfigure}{.28\textwidth}
 			\includegraphics[width=\columnwidth]{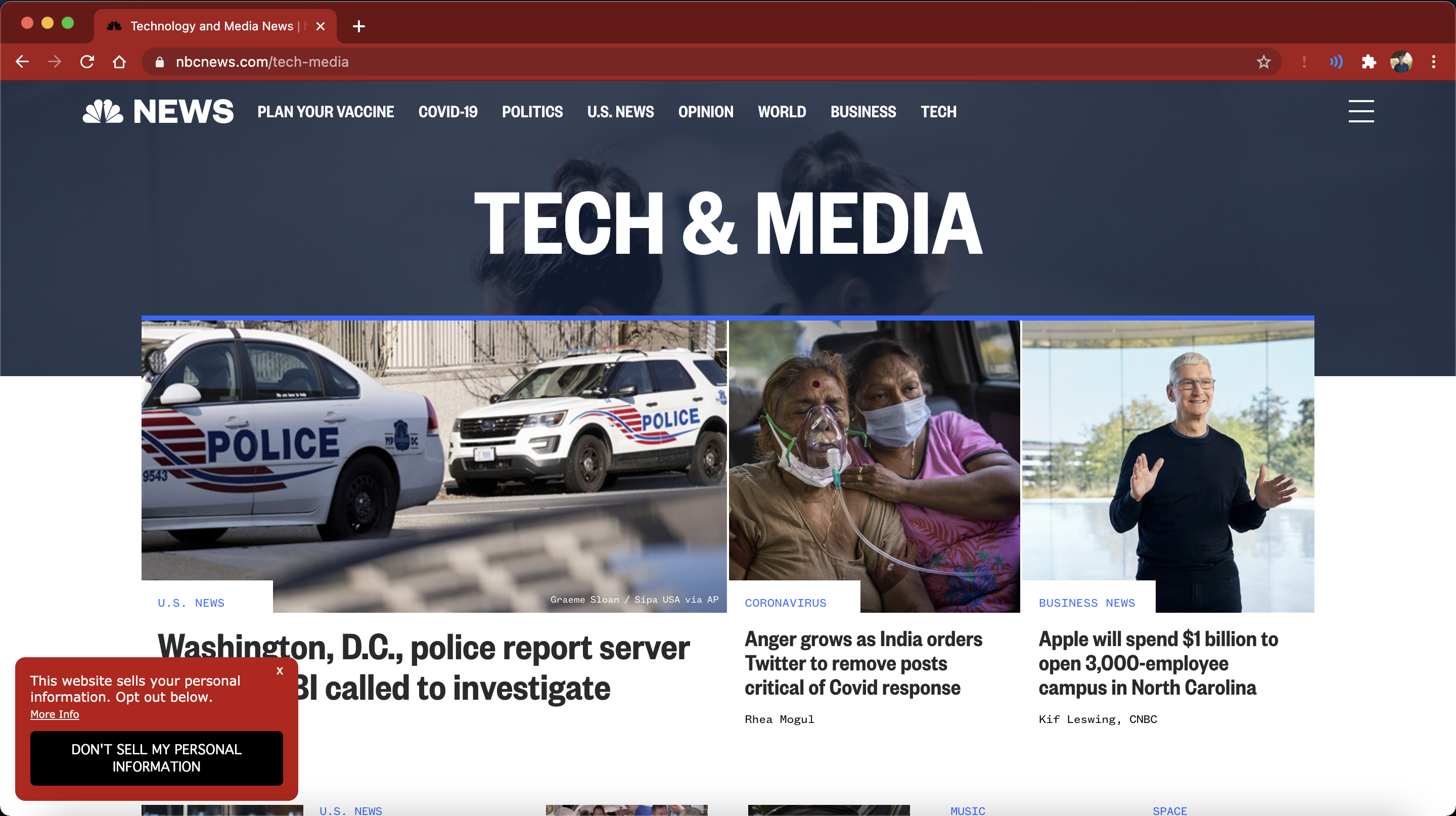}
 			\caption{Website with a COA banner.}
 			\label{fig:popup}
 		\end{subfigure}
 		\caption{Example popup banners generated by the COA browser extension. Banners appear in the bottom-left corner.}
 	\end{center}
 \end{figure*}

\section{Methodology}\label{sec:coa_design}

We designed and implemented two browser extensions: an observational extension and an opt-out assistant. We then leveraged these extensions to conduct a user study with 54 participants. 

\subsection{Observational Extension}

Our observational extension runs in the background of every webpage a user visits and searches for opt-out of sale links using a two-tiered approach. It first searches the webpage's HTML source for an opt-out of sale link with the legally-mandated label; these are considered \emph{valid} links. Since about 5\% of websites that sell data have incorrectly-labeled  opt-out links~\cite{oconnor2021clear}, the extension also searches for links with related alternate phrases (e.g., “california privacy”, “consumer privacy”, “do not sell”); these are considered \emph{ambiguous} links. We manually verified our link detection on the Alexa Top 500 websites; 
 it correctly identified all opt-out of sale links. 

To measure opt-out of sale frequency, the observational extension logs limited data about user behaviors. 
Each time a user loads a web page or clicks on an opt-out of sale link,  a log message is sent to the server. This log message contains a hash of the current website, whether or not the website contains an opt-out link (valid or ambiguous), a timestamp, and the user’s unique identifier.

\subsection{CCPA Opt-out Assistant (COA)}

COA extends the observational extension. It uses the same link detection logic as the observational extension, but it also displays a standardized banners on websites that provide opt-out of sale links. 
If COA detects a valid opt-out of sale link, the browser icon for the extension changes color to bright red, and a red popup banner (Figure~\ref{fig:popup-closeup}) appears in the bottom left corner alerting users that the website they are on sells their data. 
If COA detects an ambigious link, the browser icon for the extension will change color to yellow, and a yellow popup banner (Figure~\ref{fig:warning-closeup}) will appear in the bottom left corner. The position of these banners is depicted in Figure~\ref{fig:popup}. 

If a user clicks on the button in a COA banner, their browser is redirected to the website's opt-out page by simulating clicking on the detected HTML element in the native page (i.e., the opt-out link). 
Users who click on this button do not see the banner on subsequent visits to the same site. Users can dismiss the banner by closing it, but the banner will be re-displayed  the next time they visit the site. 


Users can also interact with COA by clicking on the browser extension icon. From the resulting menu, they can opt-out of sale on the current site or permanently dismiss the banner for this website. 

To measure the effect of visibility on opt-out of sale frequency,  COA logs limited data about user behaviors. In addition to information logged by the observational extension, COA also logs when and how the user utilizes an opt-out mechanism provided by COA (the banner or the button in the extension menu) and when a user permanently disables the banner on a site. 

\subsection{User Study}

We conducted a user study with 54 California users recruited through Amazon Mechanical Turk (MTurk). This user study was comprised of three parts:
\begin{enumerate}
	\item An initial survey about people's experience with the right to opt-out of sale. This survey asked questions about opinions and awareness of websites data sale practices and experience (or lack thereof) with CCPA opt-out of sale mechanisms.  The full set of survey questions is given in Appendix~\ref{appendix:study}. 
	\item An observational study of how people interact with opt-out of sale mechanisms. For this phase, participants were randomly assigned to one of two conditions. Half of participants were provided with instructions on how to install the observational browser extension; the other half were provided with instructions on how to install COA. All participants were instructed to keep the extension installed for at least a week, but to continue using their browser as they normally would, and interacting with the extension only as they wanted to. 
	\item A follow-up study about people's experience with the right to opt-out of sale. After one week, participants were invited to fill out a follow-up survey about people's experience with the right to opt-out of sale. This survey used the same questions as the initial survey (Appendix~\ref{appendix:study}). 
\end{enumerate}
Participation was restricted to California residents who use Google Chrome and had previously completed at least 50 HITs on MTurk with at least a 95\% acceptance rate. 
Participants who completed the first part of the study and left the browser extension installed for at least 24 hours were compensated \$5. Participants who also completed the third part of the study were paid an additional \$1.

\subsection{Ethical Considerations}

To maximize privacy, no personally-identifiable information is collected or stored by either of our browser extensions. Log records are associated with a pseudorandomly-generated identifier, and the URLs of websites visited are hashed before being sent to the server. Although hashed URLs are not fully anonymous, to ensure users' privacy we made no efforts to re-identify websites visited or analyze browsing patterns. 

Participants were informed in advance about what information would be collected and how it would be used and consented to participate in the study. These users were also informed that they had the right to opt out of the study at any time; data from users who elected to opt out would be deleted. 


This user study was reviewed in advance and granted an IRB exempt approval by the Pomona College institutional review board. 

\begin{figure*}[t!]
	\begin{subfigure}{.45\textwidth}
		\centering
		\includegraphics[width=.8\columnwidth]{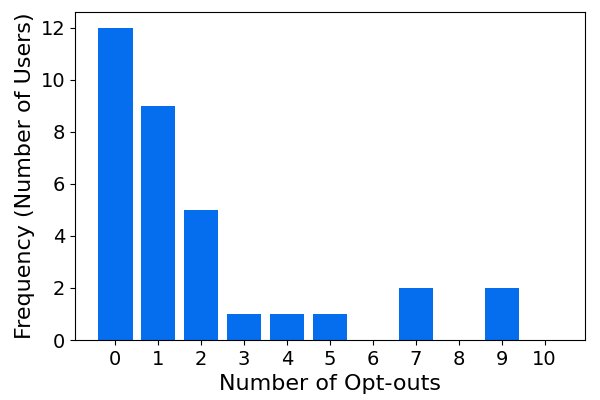}
		\caption{Frequency distribution of opt-outs with COA. 
		}\label{fig:mturk-optout-frequency}
	\end{subfigure}
	\hspace{.09\columnwidth}
\begin{subfigure}{.45\textwidth}
	\centering
	\includegraphics[width=.8\columnwidth]{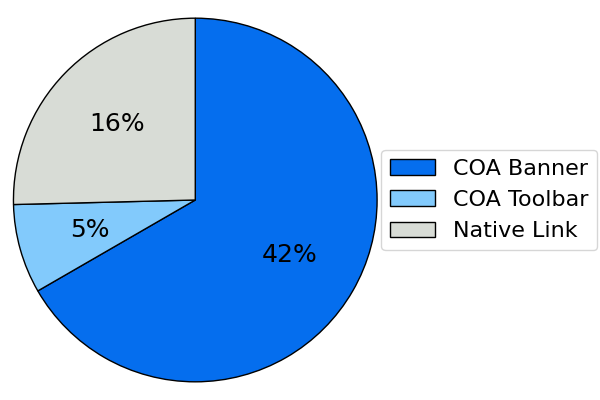}
	
	\caption{Opt-out requests from COA users by mechanism.
	}\label{fig:mturk-optout-type}
\end{subfigure}
\caption{Opt-out behavior by COA users.}

\end{figure*}

\section{Results}\label{section:results}

\begin{figure*}[t!]
	\begin{subfigure}{.45\textwidth}
		\includegraphics[width=.9\columnwidth]{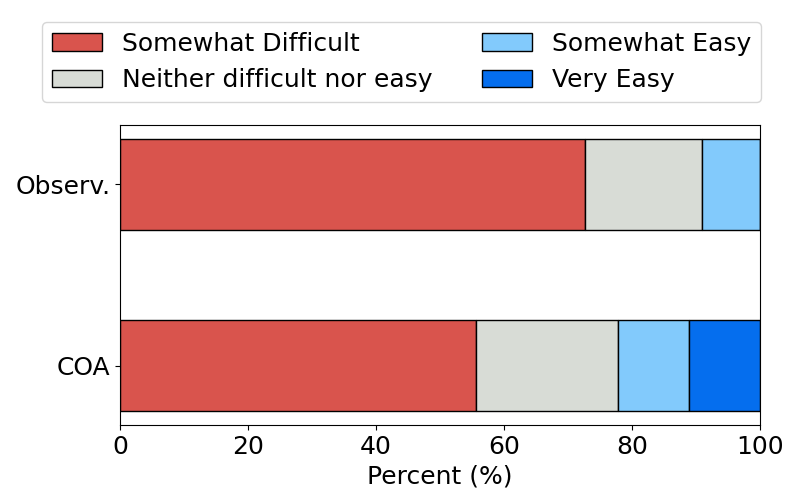}
		\caption{Difficulty of opt-out}
		\label{fig:difficulty}
	\end{subfigure}
\hspace{.09\textwidth}
	\begin{subfigure}{.45\textwidth}
		\includegraphics[width=.9\columnwidth]{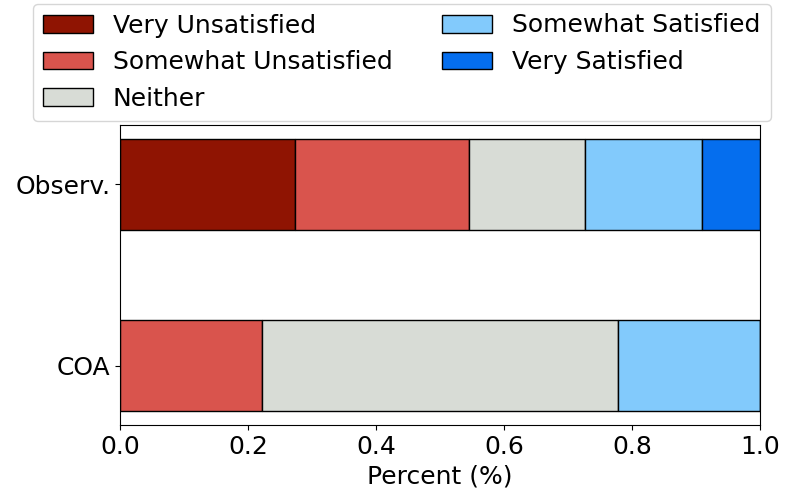}
		\caption{Satisfaction with mechanism}
		\label{fig:satisfaction}
	\end{subfigure}
	\caption{Usability of opt-out of sale features with COA and without COA. }
\end{figure*}


After eliminating users who did not install the browser extension or who uninstalled it after less than 24 hours, we had a total dataset of 54 users. 
Of these users, 32 installed and used COA and 22 installed and used the observational extension. 
24 of the participants also filled out the post-survey; half of those had used COA for a week and half had used the observational extension. Across the 54 users, we logged 51,399 total logs entries and  2,264 unique websites visited. 





\subsubsection*{Comfort with Data Sale}
In general, users were uncomfortable with the sale of their data. Only 6\% of study participants said they were somewhat or very comfortable if websites sell their personal information; a majority (54\%) said they were somewhat or very uncomfortable with the sale of their personal information.



\subsubsection*{The Effect of Visibility on Opt-out Behavior}  None of the users who installed the observational extension opted-out of sale on any websites. By contrast, 63.6\% of  participants who installed COA opted-out of sale on at least one website (Figure~\ref{fig:mturk-optout-frequency}). Overall, COA users opted out of sale on 18.8\% of sites that provided an opt-out of sale mechanism. 
This difference in opt-out frequency between observational extension users---who only had access to the native opt-out mechanisms---and COA users---who saw a banner when they had the opportunity to invoke their right to opt-out of sale---was statistically significant ($p<.001$). Moreover, the majority of opt-outs by COA users utilized the provided banners (Figure~\ref{fig:mturk-optout-type}). 
These results suggest that improving visibility of opt-out mechanisms by requiring websites to display these opportunities in banners rather than conceal them behind low-visibility links might be an effective approach to improving privacy under CCPA.

\subsubsection*{The Effect of Banners on Opt-out Usability} We compared responses on the follow-up survey between participants who used COA for a week and participants who used the observational extension with no visible features. We found that 55\% of COA users found it difficult to exercise their right to opt-out of sale; while high, this number was lower than the 72\% of  users who found it difficult to opt-out of sale without COA (Figure~\ref{fig:difficulty}). 
We also found that COA users were more satisfied with opt-out of sale mechanisms. Only 22\% of COA users reported that they were somewhat unsatisfied with opt-out of sale mechanisms and none reported that they were very unsatisfied, compared with 54\% who reported being somewhat or very unsatisfied with the opt-out of sale mechanisms provided natively by the websites (Figure~\ref{fig:satisfaction}). Persistent difficulty and low rates of  ``very satisfied'' might be due to complex and indirect mechanisms presented by websites after the opt-out link is followed~\cite{oconnor2021clear}.

\begin{figure}[t!]
	\centering
	\includegraphics[width=\columnwidth]{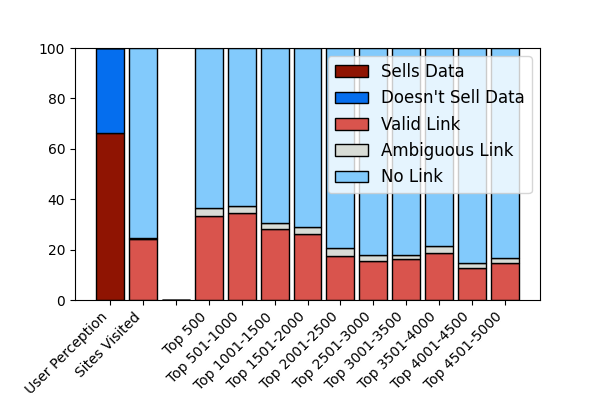}
	\caption{Perceived frequency versus actual frequency of the sale of personal information (as defined by CCPA).}
	\label{fig:sell_rates}
\end{figure}

\subsubsection*{Beliefs about Data Sale}
On average, participants estimated that 66.2\% of websites they visit sold their data. By contrast, we found that just 24.3\% of sites visited by real-world COA users contained opt-out of sale links. To better understand the prevalence of sale of personal information (as defined by CCPA),  we wrote a script that visited the Top 5000 websites (as listed by Alexa on January 12, 2021) and classified each website according to whether it had a valid opt-out of sale link, and ambiguous link, or no link. We found that 34.2\% of the Top 500 sites  and 21.7\% of the Top 5000 websites had a valid opt-out link on their homepage; an additional 3.5\% (resp. 2.7\%) had ambiguous links. In general, frequency of opt-out links decreased with site popularity.  A comparison between user perceptions about frequency of data sale and actual frequency of opt-out of sale links is given in Figure~\ref{fig:sell_rates}. While it is likely that some lower-ranked websites are not subject to CCPA requirements---the law applies only to businesses above certain revenue or user thresholds---the gap between perceptions and reality for even the top sites suggests that users significantly overestimate how common sale of personal information (as defined by CCPA) actually is. 


This over-estimate might be indicative of widespread distrust and dissatisfaction with current data practices; prior work has found that users feel helpless and frustrated about targeted advertising and the sale of their data~\cite{Mach20,Cranor12,habib2020s}. This misconception could also stem from the somewhat esoteric distinction between data sale and the broader data economy. 
Certain companies---including major data brokers like Google or Facebook---appear to perform all data processing internal to the company; such practices are not subject to CCPA's opt-out of sale requirement despite the fact that these companies benefit financially from users' personal information.  This distinction might feel unimportant to users. 

To our surprise, installing and using COA did not result in more accurate perceptions about the number of sites that sell personal information (as defined by CCPA). This might be due to a perception bias;  banner warnings on sites might have more impact on user impressions than the absence of such signals.

\section{Conclusion}


This work explores the impact of visibility on CCPA's right to opt-out of sale. We find that visible, standardized banners significantly increase opt-out rates compared to the mechanisms currently provided by websites. 
Users also found banner-based mechanisms easier to use and more satisfactory. Although banner-based consent mechanisms are clearly imperfect---prior work in the context of cookie banners has found that many users just click to get rid of banners~\cite{kulyk2020has} and that dark patterns and other UI elements can manipulate user consent decisions~\cite{nouwens2020dark,bermejo2021website,habib2022okay,ma2022prospective}---our results show that banner-based opt-out mechanisms would still significantly enhance privacy compared to the current state of the world under CCPA. 

While we have now released COA on the Chrome store, we do not consider this extension itself to be an effective tool for enhancing privacy; privacy signals are likely a better solution for users who are willing and able to proactively take steps to invoke their rights under CCPA. Instead, we view our results as evidence that effectively extending privacy rights to all users depends on regulations imposing minimum visibility standards, and that CCPA's requirement of a clear and conspicuous link on the home page falls short of this standard. Future regulations will need to provide clear and enforceable visibility requirements informed by empirical user studies in order to ensure that they actually enhance user privacy.

\bibliographystyle{plain}
\bibliography{main}

\appendix

\section{Survey Questions}\label{appendix:study}
In this Appendix, we provide the complete set of questions asked in the preliminary survey and the follow-up survey provided to study participants recruited through Amazon Mechanical Turk.

\begin{enumerate}
\item ``What percentage of the websites you visit do you believe sell your personal data?'' (Chosen on scale from 0-100)

\item ``If the websites you visited tracked your behavior and sold this information to third-parties, how comfortable would you be with it?'' (Very Comfortable / Somewhat comfortable / Neutral / Somewhat uncomfortable / Very uncomfortable)

\item ``Are you aware that California law requires websites that sell your data to allow you to opt out?'' (Yes / No)

\item ``How often have you noticed websites you visit giving you an option to opt-out of the sale of your data?'' (Never / A few times / Sometimes / Often / Always)

\item ``How often do you opt-out of the sale of your data on websites you visit?'' (Never Have / Have a few times / Sometimes / Usually / Always) \label{question:optoutfreq}

\item (If did not respond ``Never'' to Question~\ref{question:optoutfreq}) ``How difficult on average did you find it to opt-out of the sale of your data on websites you visit?'' (Somewhat difficult / Neither difficult nor easy / Somewhat easy / Very easy)

\item (If did not respond ``Never'' to Question~\ref{question:optoutfreq})  ``How satisfied are you with the mechanisms that you have used to opt-out of the sale of your data on websites you visit?''  (Very satisfied / Somewhat satisfied / Neutral / Somewhat unsatisfied / Very unsatisfied)

\item ``What is your current age?'' (18-24 / 25-34 / 35-44 / 45-59 / 60-74 / 75+)

\item ``What is your gender?'' (Man / Woman / Non-binary person / Other)

\item ``Choose one or more races that you consider yourself to be:'' (White / Black or African American / American Indian or Alaska Native / Asian / Pacific Islander or Native Hawaiian / Other)

\item ``Do you consider yourself to be Hispanic?'' (Yes / No)

\item ``In which state do you currently reside?'' (50 States / D.C. / Puerto Rico / Not in US)

\end{enumerate}

\end{document}